\documentclass[a4paper,fleqn,usenatbib]{mnras}

\usepackage{newtxtext,newtxmath}

\usepackage[T1]{fontenc}
\usepackage{ae,aecompl}

\usepackage{graphicx}	
\usepackage{amsmath}	

\title[Magnetic fluxes of solar active regions]{Magnetic fluxes of solar active regions of different magneto-morphological classes: I. Cyclic variations}

\author[V.I.Abramenko]{
Valentina I. Abramenko,\thanks{E-mail: vabramenko@gmail.com (VIA)}
Regina A. Suleymanova,
Anastasija V. Zhukova
\\
Crimean Astrophysical Observatory, Russian Academy of Science, Nauchny, Bakhchisaray,  298409, Russia
}

\date{Accepted XXX. Received YYY; in original form ZZZ}

\pubyear{2015}

\begin{document}
\label{firstpage}
\pagerange{\pageref{firstpage}--\pageref{lastpage}}
\maketitle

\begin{abstract}

Data for 3046 solar active regions (ARs) observed since May 12, 1996 to December 27, 2021 were utilized to explore how the magnetic fluxes from ARs of different complexity follow the solar cycle.  
Magnetograms from the Michelson Doppler Imager instrument on the Solar and Heliospheric Observatory and from the Helioseismic and Magnetic Imager instrument on the Solar Dynamics Observatory were utilized. Each AR was classified as a regular bipolar AR (classes A1 or A2), or as an irregular bipolar AR (class B1), or as a multipolar AR (classes B2 or B3). Unipolar ARs were segregated into a specific class U. We found the following results. 
Unsigned magnetic fluxes from ARs of different classes evolve synchronously following the cycle, the correlation coefficient between the flux curves varies in a range of (0.70 - 0.99). 
The deepest solar minimum is observed simultaneously for all classes.
Only the most simple ARs were observed during a deepest minimum: A1- and B1-class ARs.
The overall shape of a cycle is governed by the regular ARs, whereas the fine structure of a solar maximum is determined by the most complex irregular ARs. Approximately equal amount of flux (45$-$50\% of the total flux) is contributed by the A-class and B-class ARs during a solar maximum.
Thus, observations allow us to conclude that the appearance of ARs with the magnetic flux above 10$^{21}$ Mx  is caused by the solar dynamo that  operates as a unique process displaying the properties of a non-linear dynamical dissipative system with a cyclic behaviour and unavoidable fluctuations.

\end{abstract}

\begin{keywords}
Sun:magnetic fields -- Sun:photosphere -- Sun:solar cycle
\end{keywords}



\section{Introduction}

According to the Babcock$-$Leighton dynamo model \citep{Babcock1961,Leighton1969}, toroidal magnetic fields appear on the solar surface as bipolar magnetic structures, known as active regions (ARs), which obey certain rules \citep{Lidia2015}. In particular, the Hale polarity law states that the leading sunspots in bipolar ARs have a positive (negative) polarity in Northern (Southern) hemisphere in odd solar cycles, and the pattern is opposite in even cycles \citep{Hale1925}. The tilt angle of a bipolar AR usually follows the Joy's law \citep[e.g.,][]{Howard1991,Stenflo2012,Tlatova2018}. As well, a prevalence of the leading spot is usually observed in bipolar ARs \citep{Babcock1961,Lidia2015}. Bipolar ARs play a key role in the flux transport dynamo models \citep{Choudhuri1995,Dikpati2009,Charbon2010,Karak2014} as an essential agent to supply the polar zones with the poloidal magnetic flux for an oncoming cycle. Since the pioneer paper by \citet{Babcock1961}, bipolar ARs were treated as an essential type of a magnetic field configuration observed on the solar surface \citep[e.g.,][]{Lidia2015}. It would be thought that the essential part of the magnetic flux concentrated in sunspot groups also comes from regular bipolar ARs. At the same time, it is well known that the flux distribution in an AR might be very  distinct from the regular bipolar pattern \citep[e.g.,][]{Lidia2015,Toriumi2019}. Moreover, the most complicated multi-polar structures are frequently the source of the most powerful flaring activity \citep{Chen2011,Abramenko2021}. A question arises then, what part of the total AR magnetic flux can be attributed to the regular bipoles? What is the contribution from the rest of the ARs? Does the flux of different types of ARs follow a solar cycle similarly? Bipolar ARs, violating the common rules, and multipolar ARs (we will denote all of them as irregular ARs) might be a foot-print of turbulence occurring in the convection zone. Therefore the evolution of the associated flux during a solar cycle might shed light on the role of turbulence in the flux production and distortion. As a result, some information on the turbulent component of the solar dynamo can be achieved.

An aim of the present study is to address the above mentioned questions. This can be achieved by analysing a long set of magnetographic data (at least for two solar cycles), namely, independent snapshots of the solar disk. The analysed ARs should be classified as ``regular'' and ``irregular'', with a fine gradation of various deviations from a typical bipolar structure and an increasing manifestation of turbulent distortion of an emerging toroidal flux tube. 

The total flux of ARs (sunspots, or flux concentrations) was explored in many aspects. Usually, when analysing very long time series, authors use the sunspot area as a proxy for the total magnetic flux of a sunspot based on the linear relationship between the flux and the area. For example, \citet{Nag2016} utilized this approach for a 400-year long data set to explore properties of the Maunder minimum. The statistical distribution of fluxes in magnetic flux concentrations of all sizes was also investigated \citep[e.g.,][]{Hagenaar2001,Parnell2009}. Evolution of the total AR magnetic flux during its lifetime was also explored in context of the relationship between the flux and the flux emergence rate \citep[e.g.,][]{Fu2016,Norton2017,Abramenko2017,Kutsenko2019}. However, studies that would allow to address the questions posed above are not numerous. We began to address these questions in \citet{Abramenko2018} study that was based on a simplified AR classification and an incomplete data set for solar cycle 24. Here we present a follow up study using an improved classification scheme and a data set covering solar cycles 23 and 24. 
     
\section{Data and Method}

The sunspot counting technique is usually based on consideration of each AR (as a unit in a statistical ensemble), regardless of how many ARs are present on the disk at this moment. Usually, ARs are counted every day, or at the moment of their maximum development.

However, in the case of studying effectiveness of the solar dynamo in generating ARs, one has to consider how many and what kind of ARs are on the solar disk at the moment; the contribution of an AR into the total magnetic flux (from all ARs on the disk) should be estimated by introducing various measures that reflect the total magnetic flux of an AR and its complexity. Consecutive moments should be selected as independent snapshots of the solar full-disk magnetograms. Based on this idea, a technique of independent snapshots of full-disk magnetograms was suggested in \citep{Abramenko2018} and used later in \citep{Zhukova2020}. The parameters of this technique are as follows: a 9-day time sampling was adopted for compiling a set of independent magnetograms. Note that \citet{Oliver1995} showed that a sampling interval broader than 7 days ensures the independence of configurations of the sunspot area distribution over the solar visible hemisphere, in other words, in 7 days we observe a completely different pattern of the sunspots, and therefore, magnetic fluxes. So, the 9-day interval is safe in sense of ensuring the independence. Besides, the triple 9-day interval covers one Carrington rotation that was further used for data filtering. 

\begin{figure*}
\includegraphics[width=\linewidth]{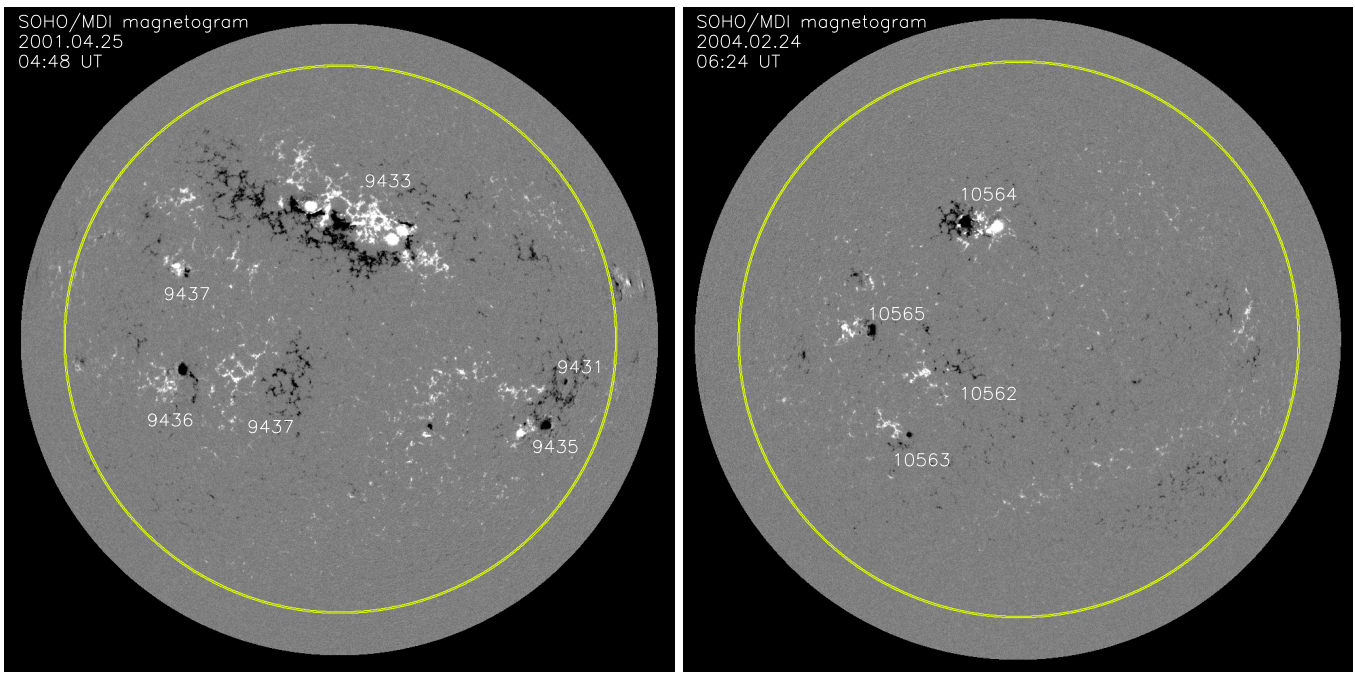}
\caption{\sf Examples of two arbitrary selected snapshots of the solar disk from the data set. Yellow circles outline a 60$^{\circ}$  wide area centered at the solar disk center. Only ARs inside this zone were counted. The left panel shows an MDI full-disk magnetogram recorded during the second peak of cycle 23 on April 25, 2001, while the right panel shows an MDI full disk magnetogram obtained on February 24, 2004 during the descending phase of cycle 23.}
\label{fig1}  
\end{figure*}

One more advantage of this choice is a convenient longitude coverage of the entire solar surface. Three consecutive  9-day-separated snapshots allow us to cut the more reliable zone of $\pm$60$^{\circ}$ from the solar disk center for further analysis of active regions inside. Figure \ref{fig1} demonstrates the 60$^{\circ}$ cutting for two full-disk magnetograms. All ARs, with the magnetic center of gravity located inside the yellow circle, were considered. 

We analysed full-disk magnetograms acquired from  May 12, 1996 to December 27, 2021. Full disk line-of-sight (LOS) magnetograms obtained before June 6, 2010 (23rd solar cycle) were recorded with the Michelson Doppler Imager (MDI) instrument on board the Solar and Heliospheric Observatory \citep[SOHO,][]{Scherrer1995} using the Ni I 6768~\AA\ spectral line with the spatial resolution of 4 arcsec.  

The 24th cycle data was acquired with the Helioseismic and Magnetic Imager (HMI) instrument on board the Solar Dynamics Observatory \citep[SDO,][]{Scherrer2012}. The magnetograms were taken at 05:24 UT in the Fe I 6173.3~\AA\ spectral line with the spatial resolution of 1\arcsec \citep{Schou2012,Liu2012}. Space-weather HMI Active Region Patches (SHARP, sharp$\_$cea$\_$720s) data, namely, LOS magnetograms and continuum images \citep{Bobra2014} were used. Details of our HMI data processing can be found in \citet{Abramenko2018}.
 
For both instruments, calculation of the total unsigned magnetic flux of an AR was performed using a proxy of the radial component of the magnetic field obtained by dividing the $B_{LOS}$-component by the cosine of the viewing angle $\mu$, the angle between the normal to the solar surface and the $LOS$ direction \citep{Hagenaar2001}.  

In total, 995 magnetograms separated by a 9 day interval were analysed, including 211 magnetograms with no ARs present on the solar disk (periods of solar minima). Two gaps in the MDI data (July 4 - October 11, 1998 and December 31, 1998 - January 27, 1999) were not considered. The total unsigned magnetic flux of an AR was summed over all pixels with the absolute value of the magnetic flux density above 50 Mx cm$^{-2}$ in case of MDI-data and above 18 Mx cm$^{-2}$ in case of HMI-data. In both cycles, only ARs with the total unsigned flux above 10$^{21}$ Mx were selected for further analysis because smaller ARs cannot be reliably identified in MDI data. 

\begin{figure*}
	\includegraphics[width=\linewidth]{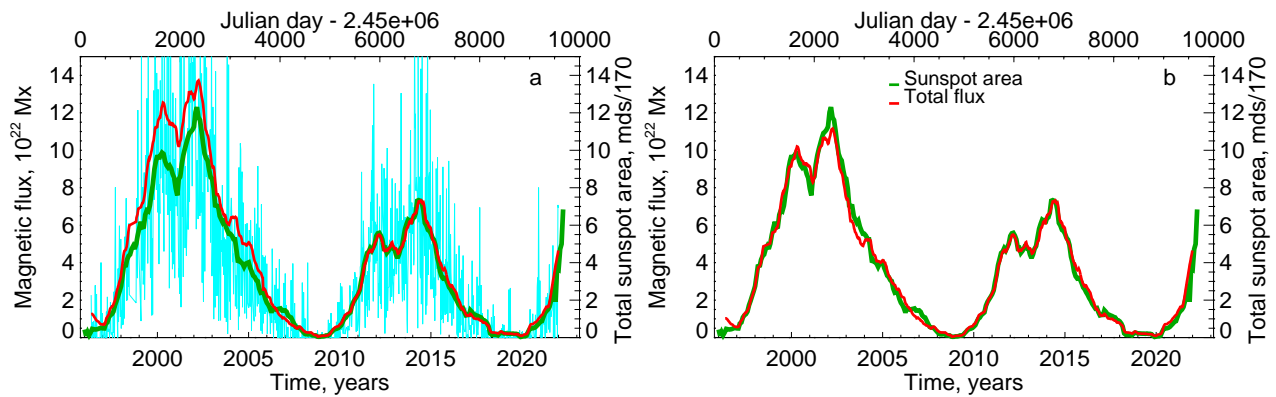}
	\caption{\sf $a$ - Monthly averages of the daily sunspot areas smoothed over 13 months (green line, right axis), magnetic flux summed over all ARs on the disc from a given-day magnetogram (turquoise line) and the same smoothed over Carrington rotation and over 13 months (red line). $b$ - the same as in panel $a$ but the cycle 23 data were preliminary divided over the MDI-to-HMI correction coefficient of 1.23 ensuring the best possible match between the area and the flux curves. 	
	}
	\label{fig2}  
\end{figure*}

It is well known that the magnetic flux measured by MDI and HMI instruments differs systematically \citep{Liu2012}. To connect properly the magnetic flux series of 23rd and 24th solar cycles, we performed the following routine. As a reference for the 23rd and 24th cycles performance, we adopted the 
USAF/NOAA monthly averages\footnote{http://solarcyclescience.com/AR-Database/sunspot-area.txt} of the daily sunspot areas (in millionths of a hemisphere). Following the traditional approach \citep[see, e.g.,][and https://www.sidc.be/silso/home]{Hathaway2015} to explore long-term variations during several solar cycles, the 13-month running mean of monthly averaged sunspot area was calculated. The resulting curve is shown with a green line in Figure \ref{fig2}. 

Magnetic flux data exceeding the above threshold were summed over the area within the 60$^{\circ}$ circle (Figure \ref{fig2}, panel a, turquoise curve) and then averaged over one Carrington rotation, after which we performed 13 month running mean averaging (Figure \ref{fig2}, panel a, red curve). While during cycle 24 the magnetic flux follows the sunspot area very well, in cycle 23 the flux curve is systematically above the area curve. It is to be expected, according to \citet{Liu2012}, who concluded that HMI magnetic flux density should be multiplied by a coefficient ($>1$) to match the corresponding values obtained with the MDI instrument. A pixel-by-pixel comparison of the magnetic flux densities were performed in \citet{Liu2012}. The authors found that the correction coefficient varies between 1.10 and 1.47 depending on the center-to-limb distance and on the magnitude of the magnetic flux density. We, therefore, accepted the correction coefficient to be 1.23 in which case we can achieve the best match between the area and the flux curves. Thus, all AR MDI data for cycle 23, were divided by 1.23, and then we repeated the calculations of the averaged flux profile (Figure \ref{fig2}, panel $b$, red curve). With the corrected data, we can be sure that the obtained magnetic fluxes, concentrated in active regions, reflect well enough the commonly accepted progress line in the two last cycles.

In total, we selected 3046 ARs. The next step of data analysis is the magneto$-$morphological classification (MMC) of ARs according to the scheme proposed in \citet{Abramenko2018}. It is based on segregation of all ARs into two basic groups which we call regular (class A) and irregular (class B) active regions.  Unipolar sunspots form a separate class U. Here, regular ARs are defined as bipolar magnetic structures that follow empirical laws (rules) \citep{Babcock1961,Lidia2015} compatible with the mean field dynamo theory, namely, the Hale polarity law, the Joy's law, and the prevalence of the leading sunspot rule. Bipolar ARs that violate at least one of these laws were assigned to class B with a specific marker to denote which rule was violated. 
This classification scheme  works well only for bipolar structures. At the same time, there is large variety of complex multipolar ARs, which produce the most powerful eruptions and solar flares. Therefore, the MMC scheme was further developed to include multi-polar ARs \citet{Abramenko2021}. Thus, A and B classes were split into A1, A2, B1, B2, and B3 sub-classes to reflect deviation of AR magnetic fields from a regular toroidal flux rope structure as the influence of turbulence in the convection zone increases: A1 type of ARs are the most regular bipoles, whereas B3 type ARs represent the most complex multipolar structures. A brief description of the classification scheme is outlined below:

\begin{description}
\item[A1 class] - 1400 ARs: Bipolar ARs obeying all criteria; no $\delta$-structures. Interpretation: emergence of a single toroidal flux tube following the global dynamo rules.
\item[A2 class] - 104 ARs: Bipolar ARs obeying all rules; a weak $\delta-$structure(s) is present. Interpretation: emergence of a single toroidal flux tube following the global dynamo weakly influenced by the turbulent component of dynamo. 
\item[B1 class] - 578 ARs: Bipolar ARs that violate at least one rule; a weak (if any) $\delta-$structure. Interpretation: result of mild distortion of a single toroidal flux tube.  
\item[B2 class] - 203 ARs: Multipolar ARs consisting of several quasi$-$coaligned bipoles having the general axis orientation in accordance with the Joy$\prime$s law; a strong  $\delta$$-$structure is frequently present. Interpretation: B2 ARs may be regarded as resulting from fragmentation and distortion of a single toroidal flux tube. A typical B2$-$class AR can be represented by NOAA AR 11158.
\item[B3 class] $-$ 180 ARs: Most complex multipolar ARs with opposite polarity sunspots distributed in an irregular manner so that it is impossible to define the axis of the AR and to assign leading and trailing sunspots. Interpretation: these ARs might result from interaction of several flux tubes in the convection zone.
\item[U1 class] - 403 ARs: Unipolar sunspots without surrounding pores and magnetic elements of opposite polarity.
\item[U2 class] - 178 ARs: Unipolar sunspots with small pores and/or magnetic elements of opposite polarity randomly distributed around the sunspot.      
\end{description}

For each AR, we determined the MMC-class during a 3-day time interval centered at the moment of the magnetogram acquisition. Note that for each MMC-class, we consider the total flux of ARs, so the number of ARs in each class does not play a role in the statistical analysis (the number of ARs in each class is not a subject for our selection,  it is a result of the dynamo action).

In the reminder of this paper, we will analyse time variations of the unsigned magnetic flux calculated separately for each MMC class.

\section{Comparison with the average solar cycle profile}
  
In spite the fact that solar cycles differ in terms of the cycle length, the peak activity level, and the rate of rise and decline of solar activity, it is still possible to generate an averaged profile of solar cycle. Following \citet{Hathaway1994,Hathaway2015}, we accept the average shape of a solar cycle (see Figure 27 in \citet{Hathaway2015}) as a ``standard'' of a solar cycle performance and we will carry out a comparison of this profile and the magnetic flux profiles generated in this study. The aim is to reveal which of the classes most closely follows the averaged cycle, in other words, how ARs of different classes comply with the ``standard'' shape of the solar cycle.   

\begin{figure*}
	\includegraphics[width=\linewidth]{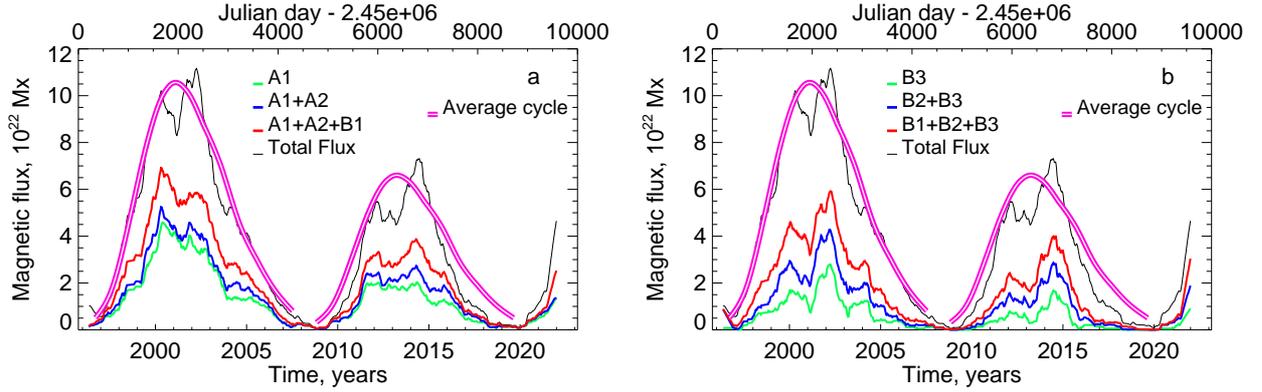}
	\caption{\sf $a$ - Time variations of the fluxes of bipolar ARs compared to the corresponding average cycle (double pink curves scaled to match the cycle amplitude). Marks from A1 to A2+A2 to A1+A2+B1 denote the flux from A1-class ARs to united flux from A1- and A2-class ARs to united flux from all bipolar ARs: A1, A2, and B1 classes. $b$ - time variations of the fluxes of irregular ARs compared to the corresponding average cycle. Marks from B3 to B2+B3 to B1+B2+B3 denote the flux from ARs of B3 class to united flux from B2- and B3-class ARs to the flux from all irregular ARs.   	  	
	}
	\label{fig3}  
\end{figure*}

\begin{table*}
	\caption{Pearson correlation coefficients, CC, between the average cycle curves and the magnetic fluxes for distinct MMC classes of ARs }
	\label{tab1}
	\begin{tabular}{ l c l c } 
		\hline
		Cycle 23    &   CC$^*$      &   Cycle 24     &    CC$^*$ 	\\
		\hline
		Total flux  &   0.980 (0.975-0.984)  &   A1           &  0.968 (0.960-0.974)	  \\
		A1+A2+B1	&	0.979 (0.974-0.983)  &	Total Flux    &	 0.967 (0.959-0.973)	  \\
		A1+A2   	&	0.974 (0.968-0.979)  &   A1+A2+B1     &	 0.967 (0.959-0.973)      \\
		A1          &   0.971 (0.964-0.976)  &   A1+A2        &  0.967 (0.959-0.973)   \\
		B1          &   0.962 (0.953-0.969)  &   B1           &  0.910 (0.889-0.927)    \\
		B1+B2+B3    &   0.956 (0.946-0.964)  &   B2           &  0.907 (0.885-0.925)   \\
		B2          &   0.918 (0.899-0.934)  &   B1+B2+B3     &  0.880 (0.852-0.903)   \\
		B2+B3       &   0.916 (0.896-0.932)  &   B2+B3        &  0.856 (0.823-0.883)   \\
		B3          &   0.869 (0.839-0.894)  &   A2           &  0.824 (0.792-0.852)   \\
	    A2          &   0.677 (0.621-0.726)  &   B3          &  0.761  (0.709-0.804)   \\
		\hline  
		\multicolumn{4}{l}{\footnotesize{$^*$ - for each cycle, the classes are arranged in the order of the CC decrease.  }}\\
		\multicolumn{4}{l}{\footnotesize{     In brackets the 0.95-confidence interval is shown.  }}\\   
		\end{tabular}	
\end{table*}

In Figure \ref{fig3} we plot magnetic flux profiles of all AR classes as well as scaled average cycle profiles (pink double line) from \citet{Hathaway2015} so that it matches the observed peak of each cycle. We see that the total observed magnetic flux (calculated from all ARs on the disk) better agrees with the average cycle for cycle 23 than it does for cycle 24, when the wings of the average cycle are shallower and broader as compared to the flux profile. This difference can also be seen in the corresponding Pearson correlation coefficients (CC, Table \ref{tab1}) showing that the correlation between the average cycle and the total flux profiles is 0.980 for cycle 23 and 0.967 for cycle 24. We note that in general the correlation is very high, which supports the underlying idea of \citet{Hathaway1994}.  

Figure \ref{fig3}, panel $a$ only shows time profiles of bipolar ARs. The flux of the most regular ARs of A1-class (green line) does not display the two-peak structure typical for the maximum phase of a solar cycle: while during cycle 23 we observe only one peak (in 2000) in the A1-profile, in cycle 24 the A1-profile is nearly flat. Adding the A2-class flux (ARs with small $\delta$-structures) resulted in the A1+A2 curve (blue) with a hint of the double-peak structure (peaks in 2002 and 2014) in both cycles. Further addition of the irregular bipoles (A1+A2+B1 red curve) reinforces this tendency, the double-peak structure is now well pronounced for both solar cycles. This experiment gives us the first hint to suggest that the two-peak shape of the solar maximum (at least, the appearance of the second sub-peak) can be due to irregularities in the magnetic configuration of ARs. The plots in panel $b$ of Figure \ref{fig3} further confirm this suggestion. Indeed, it appears that the combined contribution of all irregular B-class ARs causes the two-peak structure to appear during the solar maximum, with the secondary sub-maximum being enhanced, similar to that in the profile of the total flux. Nevertheless, the correlation with the average cycle curve is lower for the irregular ARs than that for the regular (see Table \ref{tab1}). 
 
In Table \ref{tab1}, for both cycles, the rows are arranged in the order of correlation decrease. Thus, in cycle 23 the highest correlation between the average cycle and a magnetic flux curve is observed for the total flux of all ARs observed  on the solar disk. As we proceed down along the table, we observe that the correlation becomes lower as more irregular ARs are considered. Thus, for cycle 23, the CC for the B1 class is CC=0.962 (0.910 for cycle 24), whereas the CC for the B2+B3classes is CC=0.916 for cycle 23 and 0.856 for cycle 24. The lowest correlation, in the both cycles, was found for the most complex ARs constituting the B3-class and for the A2-class ARs which represent bipolar active regions with a small $\delta$ - spot. Although the order of the classes in Table \ref{tab1} slightly differs between the cycles, the tendency of the decreasing correlation remains the same. This might imply that the general properties of a solar cycle, such as shape, intensity, and length, are guided by the global solar dynamo and are, therefore, defined by the regular bipolar ARs, whereas individual features of a cycle are rather defined by ARs violating the empirical rules, such as irregular bipoles and complex multipolars, which are thought to be influenced by the turbulent component of the solar dynamo. Moreover, no sharp transition from the global to turbulent contribution is observed, on the contrary, the most probable scenario seems to include contemporaneous operation of the global and turbulent components. Indeed, the correlation coefficients are still high for all classes, and the general enhancement of the flux from both regular and highly irregular ARs are observed as the solar cycle proceed to its maximum. The idea of the common guider for all classes is also supported by the classes behavior during the minima. Indeed, the deepest minima are observed simultaneously for all classes from 2008.1 to 2009.5 year  and from 2018.4 to 2020.2 year (see Figure \ref{fig3} and Figures \ref{fig4} and \ref{fig5} below). 

The maximal values of the (smoothed) total flux from all ARs in the maxima of the cycles are: 1.1$\cdot 10^{23}$ Mx (per instant magnetogram) in the second sub$-$peak of the cycle 23, and 7.6$\cdot 10^{22}$ Mx in the second sub$-$peak of the cycle 24.               
 
\section{Mutual correlations of different classes}

\begin{figure*}
	\includegraphics[width=\linewidth]{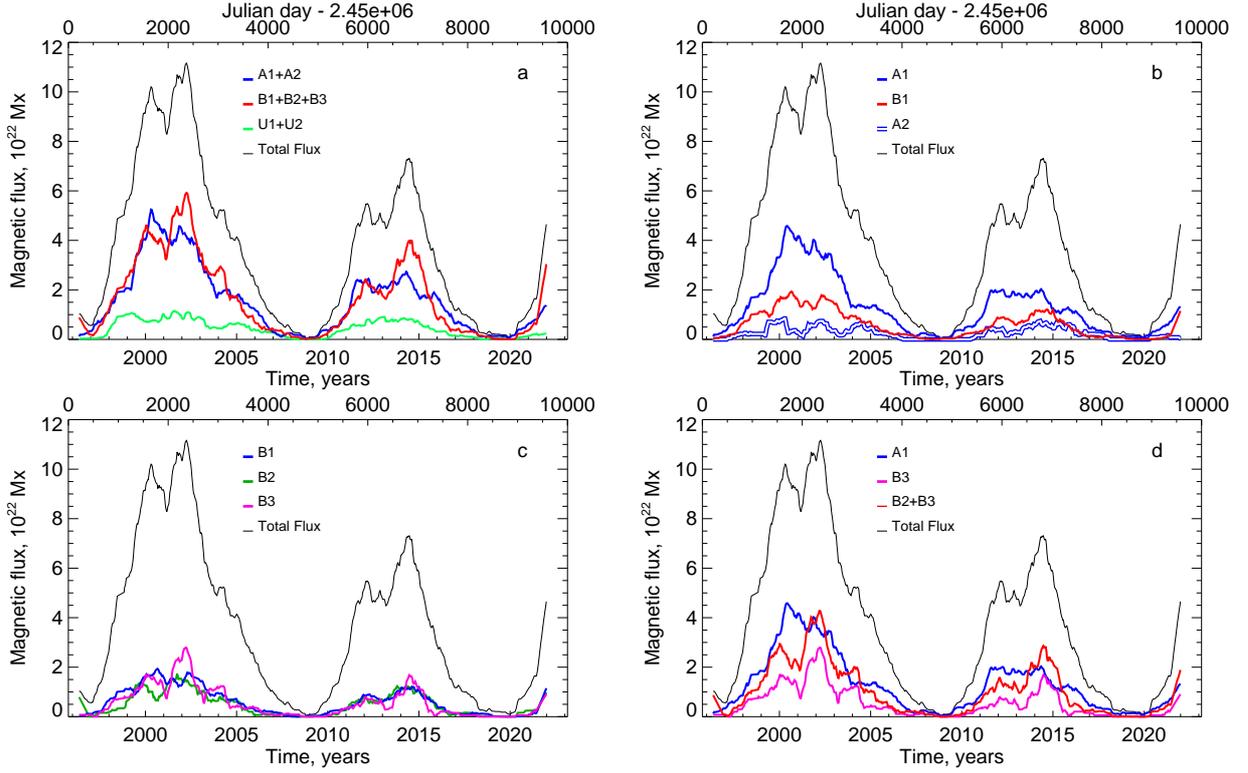}
	\caption{\sf Time variations of the fluxes of ARs of different classes. $a$ - Fluxes of all regular ARs (A1+A2), all irregular ARs (B1+B2+B3), all unipolar ARs (U1+U2).  
	$b$ - Fluxes of various bipolar ARs. $c$ - Fluxes of irregular ARs of various classes. $d$ - comparison of the most regular (A1) with the most irregular (B3 and B2+B3) AR fluxes. The correlation coefficients are shown in Table \ref{tab2}.   	   	  	
	}
	\label{fig4}  
\end{figure*}

In Figure \ref{fig4} we present flux profiles for various combinations of A and B classes and the mutual Pearson correlation coefficients are listed in Table \ref{tab2}. 

Panel $a$ in Figure \ref{fig4} plots data for three essential types of ARs that include combined fluxes from all regular ARs (A1+A2, blue curve), all irregular ARs (B1+B2+B3, red curve), and that for the unipolar ARs (U1+U2). The data show that the fluxes of regular and irregular ARs are strong and quit comparable, whereas the unipolar ARs flux is much lower and does not follow the two-peak trend seen in the other profiles. Panel $b$ in Figure \ref{fig4} shows flux profiles for various bipolar ARs. The A1 class flux appears to be the strongest, and these ARs are the most numerous (1400 ARs). We can presume that bipoles of this class determine the first maximum (2000) in the cycle 23. In the cycle 24 these ARs demonstrate a rather flat flux profile along the entire maximum (2011-2014). Anyway, in both cycles, this kind of ARs does not seem to be determining the intensity of the second maximum (2002 and 2014). The bipoles of class A2 are not numerous (104 ARs in total) and their flux is low, with wavy profile without peculiarities. The irregular bipoles (B1-class) produced a flux profile showing hints of the two-peak structure in the both cycles. We note that the combined A1+A2+B1 flux correlates very well with the total flux (see Table \ref{tab2}, second column). When moving along the column from the A1 to A1+A2 to A1+A2+B1 rows the correlation improves (with the only exception for A2 class, see discussion below). This allows us to suggest that i) the bipolar ARs are the structures that determine the overall behaviour of a solar cycle, and ii) regular and irregular bipolar ARs have the common origin. The mechanism might be the traditional toroidal flux tube rising, but with a slight rotation and/or inclination of a tube by turbulence in the convection zone in the case of an irregular bipole. As for A2-class ARs, they demonstrate unusually low correlation with all other ARs. A reason might be our unjustified assignment of this class so that these ARs are simply insignificant subset of A1 class. On the other hand, the highest correlation that they demonstrate is the correlation with B2 class - active regions consisting of several quasi$-$coaligned bipoles or large $\delta$-spots, so, A2- and B2-class ARs might have something common in topology and origin. Anyway, as soon as the flux from A2 class is very low, we prone to consider A1 and A1+A2 fluxes without a special attention to A2 flux, leaving this curious class for future investigations.     

\begin{table*}
	\caption{Pearson correlation coefficients between the magnetic fluxes of different AR-classes as derived from   the 1996-2021 data set}
	\label{tab2}
	\begin{tabular}{ l c c c c c c c c c c} 
		\hline
		            &   Total flux           &  A1 & A2  &A1+A2 &A1+A2+B1& B1  & B2 &  B3 & B2+B3 & B1+B2+B3  \\
			\hline 
		Total flux  &   1.0                  &   -  &  -   &   -    & -   &	-  &  -  &  -  &  -  &         \\
		A1       	&	0.966(2)$^*$         &	1.0  &  -   &   -    & -   &	-  &  -  &  -  &  -  &         \\
		A2          &   0.809(14) & 0.698(34)&  1.0 & -  & -  & -  & -  & -  & -  &  \\ 
		A1+A2   	&	0.982(1)  & 0.992(0)& 0.783(26) & 1.0  &   -    & -   & -  &  -  &  -  &  -           \\
		A1+A2+B1    &   0.989(1)  & 0.988(0)& 0.782(26) & 0.996(0)&   1.0  & -   & -  &  -  &  -  &  -           \\
		B1          &   0.975(1)  & 0.945(3)& 0.757(28) & 0.954(3)& 0.976(1)  & 1.0 & -  &  -  &  -  &  -           \\
		B2          &   0.938(4)  & 0.855(9)& 0.815(22) & 0.886(7)& 0.896(7)  &0.892(7)& 1.0&  -  &  -  &  - \\
		B3          &   0.935(4)  & 0.859(9)& 0.749(29) & 0.879(8)& 0.893(7)  &0.899(7)&0.879(8)& 1.0&  -  &  -\\
		B2+B3       &   0.966(2)  & 0.884(8)& 0.802(24) &0.909(6)& 0.922(5)  &0.924(5)&0.959(3)&0.978(1)&1.0&  -       \\
		B1+B2+B3    &   0.935(4)  & 0.921(5)& 0.800(24) & 0.879(8)& 0.957(3)  &0.966(2)&0.953(3)&0.968(2)&0.991(0)& 1.0         \\
		\hline  
		\multicolumn{6}{l}{\footnotesize{$^*$ - in brackets the last digit(s) of the (one-side) 0.95-confidence interval is shown.  }}\\ 
	\end{tabular}	
\end{table*}

Panel $c$ in Figure \ref{fig4} shows the profiles of irregular ARs for each class individually. The curves are quit close each other. Their combined flux (see red curve in Panel $a$) contributes in the total flux about the same amount as regular ARs do (see also a discussion in Section 5 below). 

Panel $d$ overplots fluxes of the most regular ARs (A1) and the most irregular ARs in two combinations: B2+B3 flux (all multipolar ARs) and B3 flux (the most complex multipolars). The latter two profiles undulate almost synchronously (CC=0.978$\pm$0.001), which also suggests a common formation mechanism (presumably, strong influence of the turbulent component of dynamo on a rising toroidal flux tube). The most notable feature is the clear dissimilarity between the A1 profile and B2+B3, and B3 profiles. The two-peak shape of the maxima is well pronounced in the B-profiles, but there no hint of it in the A1 profile.  The first peak seems to be caused by both, A1 and B2+B3 profiles in  both cycles, but the second peak, which is dominant in both cycles, appears only due to B-class ARs. The low correlation between the above classes further highlights the difference in the flux time profiles, showing that the A1 and B2 fluxes are correlated with CC=0.855$\pm$0.009, and A1 and B3 fluxes are correlated with CC=0.859$\pm$0.009 (see Table \ref{tab2}). One might suggest that distinct properties of the solar dynamo might be responsible for the first and second activity maxima of a cycle. Contribution of the turbulent component of solar dynamo seems to become pronounced gradually starting from mild distortions of the B1-ARs to the flux tube fragmentation for the B2-ARs to a complex intertwining of several flux tubes in the B3-ARs  (B1-ARs correlate better with the A1 and A1+A2 ARs than with the B2 and B3-ARs).

\section{Temporal Variations of Flux Ratios}

\begin{figure*}
	\includegraphics[width=\linewidth]{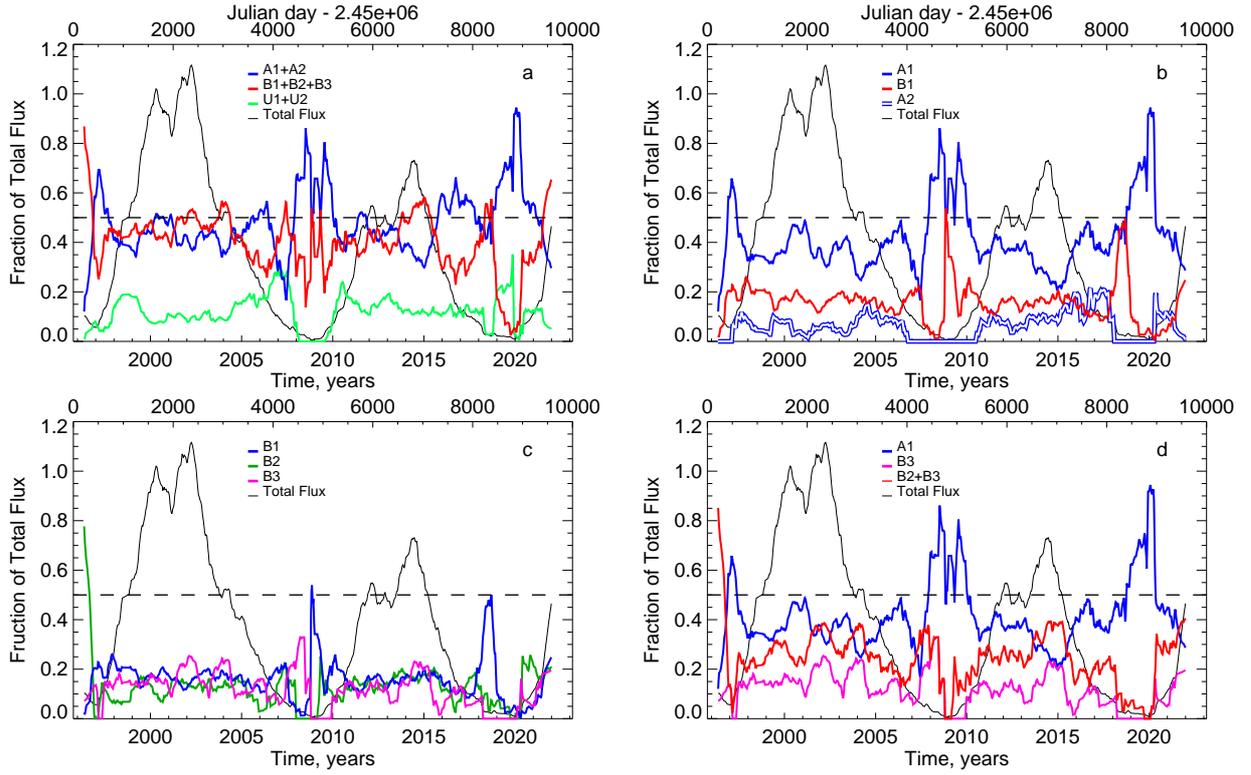}
	\caption{\sf Time variations of the fractions from the total flux of the fluxes of ARs of different classes. Notations are the same as in Figure \ref{fig4}.  	   	  	
	}
	\label{fig5}  
\end{figure*}

As it follows from Figure \ref{fig4}, ARs of distinct classes contribute differently into the total flux. Figure \ref{fig5} shows the ratios of a flux from a given class relative to the total flux. Regular (class A) and irregular (class B) ARs nearly equally contribute to the total flux, whereas the ratio of the unipolar ARs flux is very low (about 0.1). Due to the low contribution of unipolar ARs, the ratios of regular and irregular ARs vary in anti$-$phase. During the first sub-maximum of each cycle, the A and B flux ratios vary at the 0.45$\pm0.05$ level, however during the second sub-maximum in both cycles the irregular AR flux begins to dominate and the B1+B2+B3 flux ratio (see Panel $a$) reaches a value of 0.55 (with the average value of 0.50$\pm0.05$) in the 23rd and 0.57 (with the average value of 0.48$\pm0.07$) in the 24th cycle. 

So, during the first sub-maximum the flux fractions from the regular and irregular ARs nearly the same (within the error bars), whereas during the second sub-maximum the fraction of irregular ARs in 1.25-1.30 times exceeds the  flux fraction from regular ARs. 
This inference is in agreement with the finding presented  in the previous section that the appearance of the second sub-maximum is mostly caused by the magnetic flux from the irregular ARs. Panels $b$ and $c$ (purple curves) show that the enhancement is mostly due to the B3-class ARs - the most complex and the most scanty ones.       

Large and rapid fluctuations during solar minima are observed in all ratio profiles (Figure \ref{fig5}). This partially may be caused by very low flux values measured at that period, when  no A2-class ARs (regular bipolar ARs with small $\delta$-structures) were registered, while regular ARs were presented only by the A1-class ARs. Assuming that the appearance of small $\delta$-structures is due to turbulence in the upper-layer convection zone, we come to a conclusion that during a solar minimum the near$-$surface turbulence may be significantly  suppressed. Similarly, unipolar sunspots are frequently remains of large ARs, which do not appear during  a solar minimum, therefore contribution from unipolar ARs was non-existent during the 2008$-$2009 minimum.

As for the irregular ARs, only B1-class ARs (reverse polarity, wrong tilt, or dominating following sunspot) showed a well-pronounced peak during the deep minimum (see Panels $b$ and $c$ in Figure \ref{fig5}). More complex ARs (B2 and B3 classes) did not show any enhancements during the minimum (see Panels $c$ and $d$ in Figure \ref{fig5}). Note that the magnetic flux, averaged over both minima, was found to be 1.3$\pm$0.1$\cdot$10$^{21}$ Mx for regular ARs (A1-class ARs), and 0.53$\pm$0.01$\cdot$10$^{21}$ Mx for irregular ARs (predominantly B1-class ARs).

In Paned $d$ of Figure \ref{fig5} we plot fractions (of the total flux) for the most regular and the most irregular ARs, which suggest that: i) the large part of the flux measured during the first sub-maximum is contributed by the most regular A1-class ARs; ii) the flux enhancement during the second sub-maximum is due to the most irregular (B2 and B3 class) ARs, especially in the cycle 24; iii) the flux during the both solar minima is contributed predominantly by the most regular A1-class bipolar ARs.

When analyzing an ensemble of events with the aim to reveal the presence of non-linearity and self-organization, it is useful to compare the relationship between the behaviour of different statistical moments or their proxies. If non-linearity is present then higher statistical moments grow faster. In reality, the simplest manifestation of non-linearity is in the fact that a small part of specific events may play a significant or even dominant role in a process under study. 

In our case, the total number of ARs of a given class and their fluxes, expressed as fractions of the total number of ARs and the total flux, should be compared. Here the term ``total'' denotes the values summed over the entire interval of observations, i.e., over the two cycles. Results of analysis are shown in Table \ref{tab3}. 

The unipolar spots are the least flux-effective. While they constitute $p=$19\% of all ARs, their flux is only $q=$12\% of the total flux. The ratio $p/q$ for regular ARs (A1+A2) is close to unity since they represent about $p=$49\% of all ARs and contribute about $q=$43\% into the total flux. Unexpectedly, the irregular bipoles (B1-class ARs) behave similarly to the regular ARs (19\% and 16\% respectivelly). This  might suggest a very mild influence of the convection zone turbulence into the formation of irregular bipoles, namely, only the AR orientation and/or the inclination of bipole loops are affected, but the amount of flux is determined in the same way as it is for the regular bipoles (by the global toroidal field).  

In contrast, the relative contribution of complex ARs to the total flux more than twice exceeds their contribution to the total number of ARs, and the ratio $q/p$ (last column in Table \ref{tab3}) increases from B2 to B3-class ARs. This provides another evidence that B3-class ARs are more complex and more affected by non-linear influence than the B2-class ARs, and thus the proposed classification scheme is further validated. Note also that an analysis of flaring activity of ARs showed that the production of X-class flares (the strongest observed flares) is mostly caused by the B2- and B3-class ARs \citep{Abramenko2021}.  All super-active regions in the 23rd cycle \citep{Chen2011} belong to our classes B2 or B3. So, the less populated subset of ARs determine a considerable portion of the magnetic flux and is responsible for the extreme energy release capability. These are the properties of the nonlinear dynamical dissipative system of generation of the magnetic flux.

 \begin{table*}
 	\caption{Ratios of the number of ARs of a given class and of their total flux to the total number of ARs and the total flux as derived from data for the two solar cycles.}
 	\label{tab3}
 	\begin{tabular}{ l c c c} 
 	
 		\hline 
 		Class       &   $p^{a}$  &   $q^{b}$ &  $q/p$    \\
 		\hline
 		U1+U2     	&	0.191    &   0.117   & 0.612     \\
 		A1+A2   	&	0.493    &   0.433   & 0.878     \\
 		B1          &   0.190    &   0.162   & 0.853     \\
 		B2          &   0.066    &   0.136   & 2.061     \\
 		B3          &   0.059    &   0.150   & 2.542     \\
 		\hline  
 		\multicolumn{4}{l}{\footnotesize{$^a - p  -  $ Fraction from the Total number of all ARs  }}\\ 
 		\multicolumn{4}{l}{\footnotesize{$^b - q  -  $ Fraction from the Total flux from all ARs   }}\\ 
 	\end{tabular}	
 \end{table*} 

\section{Concluding remarks}

The compiled set of independent nine-day-separated full-disk solar magnetograms allowed us to explore time variations of the disk-integrated magnetic flux contributed by ARs, which were grouped following the MMC scheme: regular ARs (classes A1 and A2), irregular ARs (classes B1, B2 and B3) and unipolar sunspots (classes U1 and U2). The analysis covered two solar cycles starting May 12, 1996 until December 27, 2021. The goal of the study was to reveal how ARs of different classes follow solar cycle variations, and to infer information about the solar dynamo performance. We made the following conclusions.

\begin{description} 
\item - There exists good general correspondence between solar cycle variations and time variation of magnetic fluxes derived for each  MMC class. The Pearson correlation coefficient, CC, between a given-class flux profile and the average solar cycle profile varies in a range of  0.68-0.98, while mutual correlations between the classes is high with CC = (0.70 - 0.99). The deepest minimum in the both cycles is observed for all classes  simultaneously. This allows us to suggest that the sunspot activity is guided by the common mechanism, i.e., in generation of sunspots, the solar dynamo operates as a unique mechanism, we observe no evidence for existence of cycle-independent dynamo.  

\item - While the overall shape of a cycle is determined by regular and irregular bipolar ARs, the fine structure of the cycle profile in the period of solar maximum is mostly shaped by the most complex irregular ARs (B2 and B3 classes). The enhancement of the total flux during the second maximum in each cycle (2002 and 2014) can be assigned to the amplified flux of B2 and and B3-class ARs during that time. 

\item - The highest correlation between the total magnetic flux profile and specific class profiles was found for  regular (A1+A2) and all bipolar (A1+A2+B1) ARs. Gradual diminishing of the correlation coefficient is observed with the transition to the classes with more complex structure (with one exception of the A2 class). It is likely that contribution of additional components of the solar dynamo (for example, the turbulent component) into AR formation process increases gradually as the AR complexity increases. 
        
\item - The composition of the total (from all ARs present on the solar disk) flux was found to be the following. In each solar cycle, flux contributions from both regular and irregular ARs are nearly equal (about 45\%) during the first peak, whereas during the second peak the contribution from irregular ARs increases up to 55\% in cycle 23 and 57\% in cycle 24. At the same time, all irregular ARs constitute only 32\% of the total number of ARs. If we accept that the turbulent component of the solar dynamo affects their complex structure then we may conclude that this component is very effective: at least, during the active phase of a cycle, about a half the the magnetic flux in ARs is impacted by turbulence in the convection zone.   
 
\item - During the deepest solar minimum, only the least complex A1 and B1 ARs classes were observed. The average over both minima magnetic flux (per instant magnetogram) from regular ARs was found to be 1.3$\pm$0.1$\cdot$10$^{21}$ Mx and 0.53$\pm$0.01$\cdot$10$^{21}$ Mx for irregular ARs. 
\end{description}

In summary, we conclude that analysis of sunspot data implies that the solar dynamo operates as a unique process displaying properties of a non-linear dynamical dissipative system with a cyclic behavior. All manifestations of sunspot activity (ARs of all classes) evolve synchronously following the solar cycle. The overall shape of the cycle is governed by the subset of regular ARs (a product of emerging toroidal flux tubes), whereas the fine structures of the solar maximum are determined by the most complex irregular ARs. This implies the contributions of turbulence inside the convection zone, i.e., the existence of the turbulent component of the solar dynamo. As long as the number and the flux of the most complex ARs (B2 and B3-classes) ceases (down to zero) during a deepest minimum, we can infer that the bulk of the flux for B2 and B3 ARs  still comes from the global toroidal field and turbulent convection hardly generates much flux. Nevertheless, the presence of turbulent fluctuations is the key ingredient for solar dynamo modelling, as it follows from the 3D MHD simulations \citet[e.g.,][]{Miesch2000,Mantere2012,Fan2014}. For example, a numerical model of the AR formation \citep{Chen2017}, based on the emergence of flux bundles generated in a solar convective dynamo \citep{Fan2014} clearly demonstrates a scenario for the formation of irregular magnetic structures like those classified here as B1, B2 or B3-class ARs (see Figure 2 in \citet{Chen2017}). As for the B1-class ARs (bipole structures violating either Hale polarity law, or Joy's law, or the rule of the leading spot prevalence), in spite of the fact that we reveal nothing peculiar in sense of their magnetic flux contribution, these ARs, as it is shown by \citet{Nagy2017} ("rogue" ARs), can significantly impact  hemispheric asymmetries and even development of the current and subsequent cycles.       

The double-peak profile of a solar maximum is the widely known phenomenon \citep{Gnev1967,Gnev1977,Femi1997,Norton2010}, to mention a few. It is thought that fluctuations in the Babcock-Leighton process produce fluctuations in the polar field that are the source of fluctuations in the oncoming cycle \citep{Cameron2013,Jiang2015,Mordvinov2016,Mordvinov2019}.  As it was found recently by \citet{Karak2018}, fluctuations in the Babcock$-$Leighton process, in particular, fluctuations in the tilt angle, emergence rate, meridional circulation speed, etc., can explain the double/multi-peak profile of a solar cycle (besides, the peaks can be distributed over the entire cycle). The present research adds to this picture an explanation of the nature of the second sub-peak in the observed cycles. Our research shows that the prominent sub-peak of the irregular flux occurs during the end of the maximum, when the convection zone is already filled up by magnetic bundles of debris from failed flux emergences, submerged loops, magnetic disconnections, etc.  Convection operates as a self-organizing factor \citep{Aschwanden2018} forming concentrations and channels to rise up for these complex knots of intertwined magnetic bundles. The process ends up by the enhanced flux concentrated in irregular ARs. 

It is worth to note that certain inferences of the present study are in a good qualitative agreement with the previously published results based on a statistical analysis of the ARs number and on other classification schemes. In particular, \citet{Kilcik2011} using the McIntosh classification demonstrated that in cycle 23 large (complex) solar groups showed a two-peak shape of the group number with a maximum in 2002, whereas small (simple) groups peaked two years earlier. As a whole, basing on results for four cycles (20-23), Authors reported that the large groups number peaks on the half-way of the cycle (cycle phases 0.45-0.5), whereas ISSN usually peaks at the phase 0.29-0.35.  \citet{Jaeggli2016} using the Mount Wilson classification for cycles 23 and 24 found that the highest number of complex ARs was reached the same year as the total number of ARs, however the number of simple ARs peaked 1-2 years earlier. The observed increase of the complex ARs population during the maximum and declining phase was explained by "the collision of separate systems of the emerging flux" \citep{Jaeggli2016}. Authors also conclude that "complex ARs and simple ARs show no significant difference in distribution with latitude during the solar cycle, implying that they originate from the same reservoir of flux in the solar interior."
The two-year delay of the maximum number of complex ARs relative to the maximum number of simple ARs was confirmed in \citet{Nik2019}. Analysing the latitudinal distribution of simple and complex ARs in N- and S-hemispheres, authors concluded that simple and complex ARs belong to the same population, and therefore, they "have the same origin in the solar interior." The inference agrees with \citep{Jaeggli2016} and with our above conclusion that the solar dynamo operates as a unique mechanism in production of active regions. 

The above mentioned researchers of the simple and complex ARs populations agree in question about the fraction of the complex ARs:  16\% and 12\% was reported in \citet{Jaeggli2016}  and \citet{Nik2019}, respectively. Our counterpart for "complex" ARs is B2- and B3-class ARs, which consists 13\% from the total number of ARs (see Table \ref{tab3}). Results in \citet{Kilcik2011} qualitatively comply with these inferences: Figure 3 in  \citet{Kilcik2011} allows to infer that the mean daily sunspot group number for large ARs is systematically lower than that for small ARs for all analysed cycles. 
What new is delivered by studying magnetic fluxes of ARs is their significant contribution into the total flux, in spite of their modest population.  

\section*{Acknowledgements}

Authors are thankful to anonymous referee for productive discussion and very useful comments which helped to improve the article. 
SDO is a mission for NASA Living With a Star (LWS) program. The SDO/HMI data were provided by the Joint Science Operation Center (JSOC). The study was supported by Russian Science Foundation grant 18-12-00131.

\section*{Data availability}

The MDI and HMI data that support the findings of this study are available
in the JSOC (http://jsoc.stanford.edu/) and can be accessed under
open for all data policy.

Derived data products supporting the findings of this study
are available from the corresponding author VA on request.







\bsp	
\label{lastpage}
\end{document}